\documentclass[preprint]{elsarticle}

\usepackage{graphicx}
\usepackage{amssymb}
\usepackage{bm}
\usepackage{amsmath}
\usepackage{color,soul}

\journal{Localisation2020 - Special Issue in Annals of Physics}

\begin{document}
\begin{frontmatter}

\title{Non-local corrections to the typical medium theory of Anderson localization}


\author[1] {H. Terletska}
\author[1] {A. Moilanen}
\author[2,3]{K.-M. Tam}
\author[4] {Y. Zhang}
\author[6]{Y. Wang}
\author[7]{M. Eisenbach}
\author[8] {N. S. Vidhyadhiraja}
\author[9]{L. Chioncel}
\author[2,3] {J. Moreno}

\address[1]{Department of Physics \& Astronomy, Computational Science Program, Middle Tennessee State University, Murfreesboro, Tennessee 37132, USA}

\address[2]{Department of Physics \& Astronomy, Louisiana State University, Baton Rouge, Louisiana 70803, USA}
\address[3]{Center for Computation \& Technology, Louisiana State University, Baton Rouge, LA 70803, USA}

\address[4] {Kavli Institute for Theoretical Sciences, University of the Chinese Academy of Sciences, Beijing, 100190, China}

\address[5] {Center for Nanophase Materials Sciences, Oak Ridge National Laboratory, Oak Ridge, TN 37831, USA}

\address[6]{Pittsburgh Supercomputing Center, Carnegie Mellon University, PA 15213, USA}

\address[7]{Center for Computational Sciences, Oak Ridge National Laboratory, Oak Ridge, TN 37831, USA}

\address[8]{Theoretical Sciences Unit, Jawaharlal Nehru Centre for Advanced Scientific Research, Bangalore, India.}
\address[9]{Theoretical Physics III, Center for Electronic Correlations and Magnetism, Institute of Physics, University of Augsburg, and Augsburg Center for Innovative Technologies, University of Augsburg, D-86135 Augsburg, Germany}
%

\begin{abstract}
We use the recently developed finite cluster typical medium approach to study the Anderson localization transition in three dimensions. Applying our method to the box and binary alloy disorder distributions, we find a fast convergence with the cluster size. We demonstrate the importance of the typical medium environment and the non-local spatial correlations for the proper characterization of the localization transition. As the cluster size increases, our typical medium cluster method recovers the correct critical disorder strength for the transition. Our findings highlight the importance of the non-local cluster corrections for capturing the localization behavior of the mobility edge trajectories. 
Our results demonstrate that the typical medium cluster approach developed here provides a consistent and systematic description of the Anderson localization transition in the framework of the effective medium embedding schemes.

\end{abstract}


\end{frontmatter}



\section{Introduction}
\label{S:1}
Disorder as a ubiquitous feature of materials can cause profound effects on a variety of their properties \cite{Kramer_MacKinnon_1993,e_abrahams_10}. 
Consequently, a careful control on the concentration of defects in materials can be used to rationally change and design new functionalities of modern quantum systems. 
One of the most pronounced effects of disorder is the electron localization (Anderson localization) and the associated metal-insulator transition~\cite{p_anderson_58}. The theory of Anderson localization, where multiple scattering off impurities leads to the spatial confinement of electrons, is well developed.  It has been demonstrated that in one and two dimensions, an arbitrarily small amount of disorder localizes electrons, whereas in three dimensions states may be localized or extended depending on the amount of disorder ~\cite{e_abrahams_79}.

%
Numerical methods have played an important role in understanding the 
mechanism of Anderson localization. Several standard computational tools have been employed for finite-size lattice calculations, including exact diagonalization, the transfer matrix method, the kernel-polynomial method, as well as the multifractal approach~\cite{Kramer_MacKinnon_1993,Markos_2006,brandes2003anderson}. 
While numerically robust, the application of these methods to real materials often faces the challenge of having to treat large localization lengths when being restricted to finite system sizes. Moreover, often the techniques developed for detecting Anderson localized states, in non-interacting systems, are not directly applicable to interacting electron systems, as they are built on the knowledge of single and not many-particle eigenstates. 

Effective medium embedding methods have been developed over the past several decades; presently, these aproaches constitute an alternative and complementary way for treating disorder in materials. The most commonly used approach for disordered systems is the coherent potential approximation (CPA)  
\cite{p_soven_67,shiba71}. The CPA shares a similar conceptual construction 
with the dynamical mean-field theory (DMFT), 
which has proven to be a very successful theory for strongly correlated electron systems 
\cite{DMFT_RMP_1996,w_metzner_89a}. Both CPA and DMFT are Green's function-based methods and can be easily combined to study the interplay of disorder and electron localization. Conceptually, in these methods, the original lattice is mapped to a single impurity embedded in a  dynamical effective medium determined self-consistently. The dynamics of the medium allows for effectively capturing the disorder or interaction-induced correlations effects at the impurity level. By construction, the CPA is a local approximation, and to capture the multi-impurity scattering effects, various cluster extensions have been developed. This includes the momentum-space based Dynamical Cluster Approximation (DCA) and the real-space-cluster molecular CPA \cite{DCA_RMP,m_hettler_98a,m_jarrell_01a}.

Although these commonly used effective medium embedding methods allow studying  disorder effects, they, however, fail to capture the  Anderson localization. 
The main challenge here is that the linearly averaged (arithmetic average) density of states (DOS), obtained from the corresponding disorder averaged impurity (cluster) Green's function calculated within 
the CPA (DCA), is not critical at the Anderson transition. Hence, it cannot be used as an order parameter to characterize the localized states due to disorder.

It is therefore of interest to explore the effective medium methods that employ a proper order parameter capable of describing the Anderson localization. 
There have been several proposals according to which the Anderson transition might be detected by studying the statistical properties of the local density of states (LDOS) and its distribution \cite{p_anderson_58,g_schubert_10,m_janssen_98,Logan_Wolynes_1987,Ostlin_etal_2020,Semmler_2010,Wortis_2011, Fehske_2005}.
Dobrosavljevic et al.~\cite{v_dobrosavljevic_03} incorporated such ideas in the context of the effective medium approach. They developed the typical medium theory (TMT) and showed that the geometrically averaged LDOS is indeed an order parameter for the Anderson transition.  In the typical medium analysis, instead of using the arithmetically averaged disorder Green's function  (as it is implemented in the  CPA and the DCA), the geometrical averaging is used in the self-consistency loop. Such typical medium analysis has also been extended to interacting disordered systems \cite{k_byczuk_05,c_ekuma_15c,Dobrosavljevic_Kotliar_1997,k_byczuk_10,Byczuk_etal_2009,k_byczuk_09, Dobrosavljevic_int_2009,Dobrosavljevic_int_2013,Dobrosavljevic_int_2014,Dobrosavljevic_int_2015,Dobrosavljevic_tmt_qcp_2015,Dobrosavljevic_review_2010}. 
However, by construction, the TMT is a local single-site approximation and, hence, it neglects the non-local spatial correlations. As a result, the TMT underestimates the critical disorder strength for the Anderson transition in a three dimensional (3D) model, and does not capture properly the mobility edge trajectories for the uniform box distribution. 

To overcome such limitations, recently, we have developed the typical medium dynamical cluster approach (TMDCA) \cite{c_ekuma_14b,Terletska_etal_2018,h_terletska_14a,c_ekuma_15b,y_zhang_15a,y_zhang_16,s_sen_16a}, which is a cluster extension of the single-site typical medium
method~\cite{v_dobrosavljevic_03}. As we demonstrated in ref.~\cite{Terletska_etal_2018,c_ekuma_14a}, such typical medium analysis can properly capture the non-self-averaging behavior of the Anderson localization phenomena. In particular, it captures the dramatic changes in the distribution of the local density of states (LDOS) through the transition. At small disorder strength, the LDOS follows a Gaussian distribution, while it is a
skewed log-normal distribution at large disorder.
 The calculated typical-medium DOS (TDOS) obtained from the geometrically averaged cluster Green's function can capture such behavior. As shown in~\cite{Terletska_etal_2018,c_ekuma_14a}, the TDOS vanishes for the localized states, while it is finite for the extended states. The cluster TMDCA method when applied to the three dimensional Anderson model not only captures accurately the critical strength of the disorder, but also the reentrance behavior of the mobility edge.  
We have also extended the application of the TMDCA method beyond the simple Anderson 
model, including systems with electronic interactions~\cite{c_ekuma_15c,s_sen_16a,Sen_etal_2018}, off-diagonal disorder~\cite{h_terletska_14a,s_sen_15}, multi-band~\cite{y_zhang_15a}, and phonon systems~\cite{w_mondal_16,Mondal_etal_2019,Mondal_2020}. Recently, such developments have been successfully applied in the context of ab-initio calculations of Anderson localization in superconductors~\cite{y_zhang_15a}, dilute magnetic  semiconductors \cite{y_zhang_16}, photovoltaics \cite{Zhang_2018}, and binary alloy systems~\cite{Ostlin_etal_2020}.

The goal of the present work is to further benchmark the TMDCA method for the Anderson model in three dimensions. We perform a careful systematic cluster size analysis of the electron localization for both box and binary disorder distributions. 
Our results indicate that non-local cluster corrections are significant in capturing electron localization, hence, the finite cluster TMDCA analysis is necessary for the proper description of disorder effects in the Anderson model. 

The paper is organized as follows. 
In section~\ref{sec: method}, we provide a short  overview of the model and the TMDCA method. In section~\ref{sec: results}, we present the results of the application of the TMDCA method for the 3D Anderson model with box and binary alloy disorder distributions. We conclude, with a discussion of the prospects of the method, in the last section.

\section{Model and Method}
\label{sec: method}
We study the Anderson model of non-interacting  electrons subjected to a disordered random potential
\begin{equation}
    H = -t \sum_{<{i,j}>} (c^{\dagger}_{i} c_{j} + H.c.) + \sum_{i} V_{i} n_{i},
\end{equation}
here the operators $c^{\dagger}_{i}$ and $c_{i}$ are the creation and annihilation operators,
respectively, for an electron on site $i$, $n_{i}=c_i^{\dagger}c_i$ is the number operator, and $t$ is the hopping energy between nearest neighbors $i$ and $j$. 
The first term is the kinetic energy operator due to the hopping of electrons on a lattice, and the second term is the local on-site disorder potential.
The disorder potential is a random quantity distributed according to some specified probability distributions $P(V_i)$. We set $4t=1$ to serve as the energy scale. 

We perform our analysis for two types of disorder distribution: the (uniform) box disorder, which is given by the distribution function $P(V_i) = \frac{1}{W}\Theta(|W/2-V_i|)$ (the disorder strength is characterized by $W$), and the binary alloy disorder distribution with $P(V_i)=c_a\delta(V_i-V_A)+c_b\delta(V_i-V_b)$. Here $c_a$ is the concentration of the host $A$ ions, and $c_b=1-c_a$ stands for the concentration of the impurity $B$ ions. We introduce a shorthand notation for disorder averaged quantities: $<...>=\int dV_i P(V_i)(...)$.

The TMDCA is a typical medium extension of the conventional DCA scheme \cite{DCA_RMP,m_hettler_98a}. Just as in the DCA approach \cite{m_hettler_98a}, we map the original lattice into a cluster of size $N_c$ (constructed in momentum space) embedded in the effective medium. The effective medium is determined self-consistently. The TMDCA utilizes the geometric averaging over disorder for the cluster Green's function, while the DCA uses the arithmetic (linear) averaging. To construct a $N_c$ cluster, the first Brillouin zone is divided into $N_c=L_c^D$ ($L_c$ is the linear cluster size, $D$ is the dimension) coarse-grained cells with the cluster centers $K$ surrounded by points $\tilde{k}$ within the cell such that the lattice momentum $k=K+\tilde{k}$. Both DCA and TMDCA systematically incorporate the non-local spatial fluctuations as the cluster size $N_c$ increases, and becomes exact in the limit $N_c \rightarrow \infty$. The non-local  short-range spatial correlations are treated explicitly within the range of the cluster $N_c$, while the long length scale correlations are treated within the typical medium. As in the DCA scheme, the TMDCA self-consistency loop is constructed for the momentum $K$ dependent quantities, while to solve the cluster problem, one employs the Fourier transform to the real space of $N_c$ sites with $(I,J)$ the site
indices~\cite{DCA_RMP}.

In the TMDCA scheme, the main quantity of interest is 
the cluster typical Green's function $G_{typ}^c(K,w)$ which is obtained from the Hilbert transform of the corresponding cluster typical density of states $\rho_{typ}^{c}(K,w)$. Here $\rho_{typ}^{c}(K,w)$ is obtained using the geometrical averaging ansatz of the form \cite{Terletska_etal_2018,c_ekuma_14a}:
\begin{equation}
\rho_{typ}^c(K,w)=
\overset{local-TDOS}{{\overbrace{exp \left ( \frac{1}{N_c}\sum_{I=1}^{N_c}<\ln(\rho_I^c(w,V)> \right )}}} 
\overset{non-local}{\times {\overbrace{\left < \frac{\rho^c(K,w,V)}{1/N_c\sum_I\rho_I^c(w,V)} \right >}}}.
\label{eq: ansatz}
\end{equation}
Here $\rho_I^c(w,V)=-\frac{1}{\pi}ImG^c_{II}(w,V)$ is the local density of states at site $I$ obtained from the cluster Green's function $G^c_{IJ}(w,V)$;  $\rho^c(K,w,V)=-\frac{1}{\pi}G^c(K,w,V)$ is a non-local density of states determined from the Fourier transform of the cluster Green's function $G_{IJ}^c$. 
In the ansatz of Eq.~\ref{eq: ansatz}, to avoid self-averaging at strong disorder, we separate the "local-TDOS", which utilizes the geometric averaging over disorder, from the "non-local" $K$-dependent contributions~\cite{Terletska_etal_2018,c_ekuma_15b,c_ekuma_14b}. Later we will show that such ansatz indeed can capture effectively the electron localization in the Anderson model. To understand better the contribution to the TDOS coming from the local and non-local parts in the above ansatz, we will also perform our calculations using the "local" ansatz only, with
\begin{equation}
  \rho^{local-TDOS}_{typ}(K,w)={\exp \left ( \frac{1}{N_c}\sum_{I=1}^{N_c}<\ln(\rho_I^c(w,V)>. \right )}
  \label{eq:local-ansatz}
\end{equation}

Also notice that $\rho_{typ}^{c}(K,w)$ in Eq.~\ref{eq: ansatz} possesses the following properties~\cite{Terletska_etal_2018,c_ekuma_15b,c_ekuma_14b}: for the $N_c=1$ case, it reduces to the local TMT  with $\rho_{typ}^c(K,w)=\exp(<\ln \rho ^c(w,V)>)$. And, at weak disorder strength, the TMDCA reduces to the DCA with $\rho_{typ}^c(K,w)\rightarrow <\rho^c(K,w,V)>$.

In the following we outline the TMDCA self-consistent iterative procedure that we use in our calculations: 

1. Starting from a guess for the effective medium hybridization function $\Delta(K,w)$, we first construct the cluster-excluded Green's function 
\begin{equation}
\mathcal{G}(K,w)=\frac{1}{w-\Delta(K,w)-\bar{\epsilon}(K)},
\end{equation}
where $\bar{\epsilon}(K)$ is the coarse-grained bare dispersion. For the 3D cubic lattice, the bare lattice dispersion is given as $\varepsilon(k)=-2t(\cos(k_x)+\cos(k_y)+\cos(k_z))$. \\ 

2. Since the cluster problem is solved numerically in real space~\cite{DCA_RMP,m_hettler_98a}, we then Fourier transform $\mathcal{G}(K,w)$ to real space with $\mathcal{G}_{I,J}=\sum_K\mathcal{G}(K)e^{iK(R_I-R_J)}$.\\

3. Now we are ready to solve the cluster problem using, e.g., a random sampling. For this, we stochastically generate a random configuration of disorder potentials, 
$V_i$, and construct the cluster Green's function $G_c$ by inverting the matrix
\begin{equation}
    \hat{G}_c(V)=(\mathcal{\hat{G}}^{-1}-\hat{V})^{-1}.
\end{equation}
Then we calculate the disorder-averaged cluster typical density of states $\rho_{typ}^c(K,w)$ using the ansatz of Eq.~\ref{eq: ansatz}, and the Hilbert transform to obtain the cluster typical (geometrically averaged over disorder) Green's function 
\begin{equation}
    G_{typ}^c(K,w)=\int dw'\frac{\rho_{typ}^c(K,w')}{w-w'}
\end{equation}
\\

4. With the cluster problem solved, we close the self-consistency loop by calculating the lattice coarse-grained Green's function 
\begin{equation}
    \bar{G}(K,w)=\int \frac{N_o^c(K,\epsilon) d\epsilon}{(G_{typ}^c(K,w))^{-1}+\Delta(K,w)-\epsilon+\bar{\epsilon}(K)}
\end{equation}
which is then used to obtain a new estimate for the cluster-excluded Green's function $\mathcal{G}(K,w)$. Such an iterative procedure is repeated, until the self-consistency is reached, i.e., when the cluster typical Green's function $G_{typ}^c(K,w)$ and the coarse-grained lattice Green's function $\bar{G}(K,w)$ become equal.


\section{Results and Discussion}
\label{sec: results}
\subsection{Box disorder distribution}

\begin{figure}[htb]
\includegraphics[width=1\textwidth]{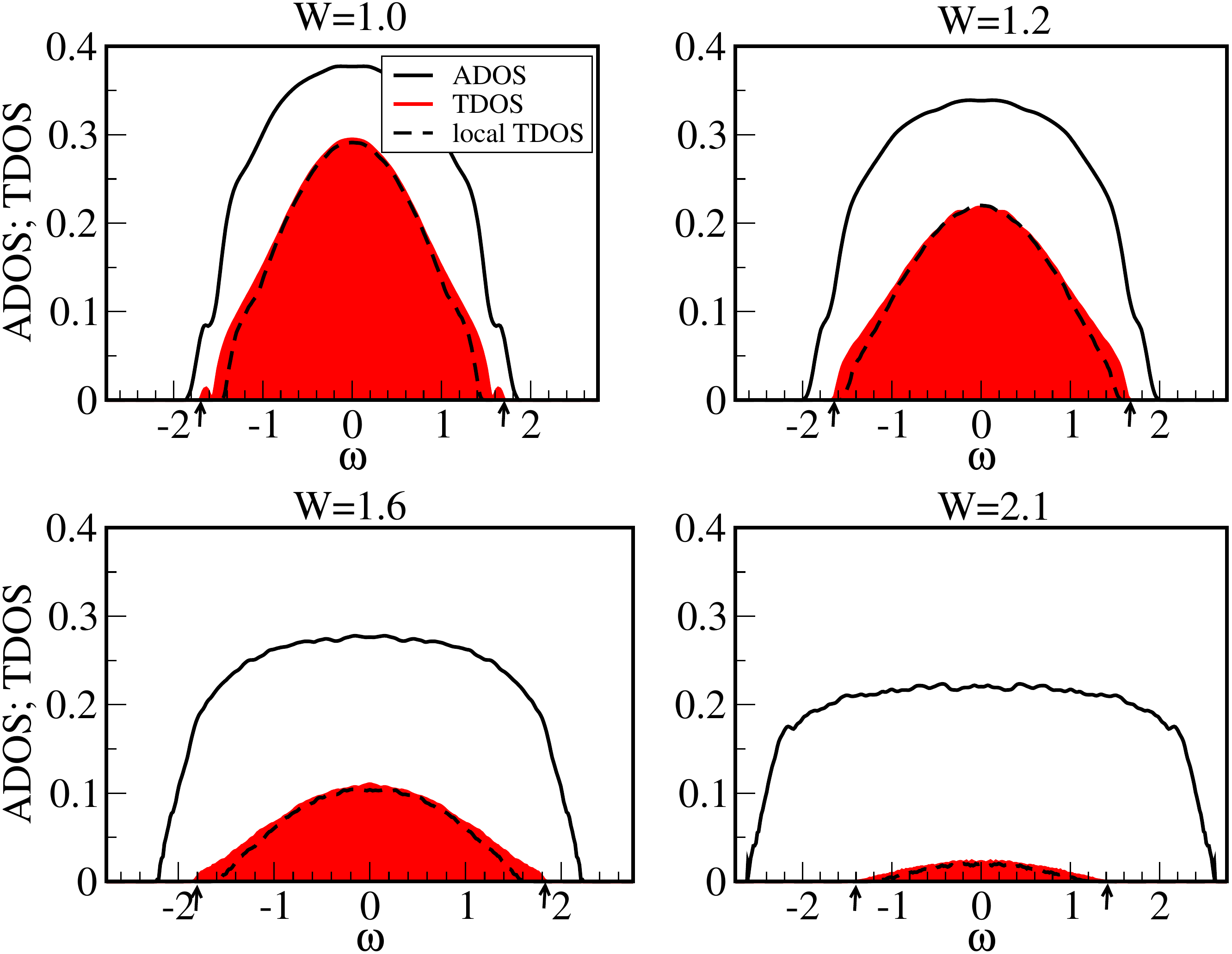}
\caption[]{ADOS (solid line) and the TDOS (shaded area) as function of frequency $\omega$ at different values of box disorder strength $W=1.0, 1.2, 1.6, 2.1$ calculated using the $N_c=64$ DCA and TMDCA methods, respectively. The local TDOS (dashed line) is obtained using Eq.~\ref{eq: ansatz}. Vertical arrows indicate the approximate position of the mobility edge boundaries. }
\label{fig:dos-box}
\end{figure}

While the localization properties of the Anderson model with box disorder distribution are well known from the literature \cite{Kramer_MacKinnon_1993}, we consider it here to demonstrate the validity of our numerical method. 
First, we start the discussion of the results by comparing the disorder evolution of the $ADOS(w)$ (obtained using the conventional DCA scheme with the arithmetic averaging over disorder in the self-consistency loop) and the typical $TDOS(w)$ (obtained from our TMDCA procedure with the geometric averaging over disorder). Our results for the cubic cluster $N_c=64$ are shown in Fig.~\ref{fig:dos-box}. For the $TDOS(w)$ , we also show the results obtained with the local ansatz of Eq.~\ref{eq:local-ansatz}.  As expected, the disorder dependence of the $ADOS(w)$ and the $TDOS(w)$ are very different: while the ADOS remains finite with increasing disorder strength, the TDOS continuously gets narrower and eventually gets fully suppressed. At weaker disorder strength $W$, the localization of electrons starts at the band tails, and is detected by vanishing $TDOS(w)$  at higher frequencies $w$. The mobility edge (shown by vertical arrows) separates the extended states (with finite TDOS) from the localized (with zero TDOS) states. As disorder strength $W$ increases, the TDOS gets suppressed at all frequencies, indicating the localization of all electrons in the band. Such suppression of the TDOS with disorder strength $W$ indicates that the TDOS indeed can serve as an order parameter for Anderson localization. 

To better understand the role of the non-local contribution (with the full momentum $K$ dependence) in the ansatz for the geometrically averaged cluster Green's function in Eq.~\ref{eq: ansatz}, in Fig.~\ref{fig:dos-box}, we also show the results for the $TDOS(w)$  obtained with the local ansatz of Eq.~\ref{eq:local-ansatz}. 
Our data indicate that the most contribution to the $TDOS(w)$ is actually coming from the geometrically averaged local DOS part in the ansatz, i.e., it is well-captured by Eq.~\ref{eq:local-ansatz}. The critical behavior at the Fermi level is the same for both the local and non-local ansatze. However, the non-local $K$- dependent contribution of Eq.~\ref{eq: ansatz} seems to be important for capturing properly the mobility edge behavior (marked by vertical arrows). Here, at edges, we observe the biggest difference between the local $TDOS(w)$  and the $TDOS(w)$  obtained with full $K$ momentum dependence. This indicates that while the critical behavior at the band center is captured properly in the local ansatz with the geometric averaging over the LDOS, the mobility edge trajectories for the local ansatz case will, however, converge slower with the cluster size $N_c$. 

\begin{figure}[htb]
\includegraphics[width=1\textwidth]{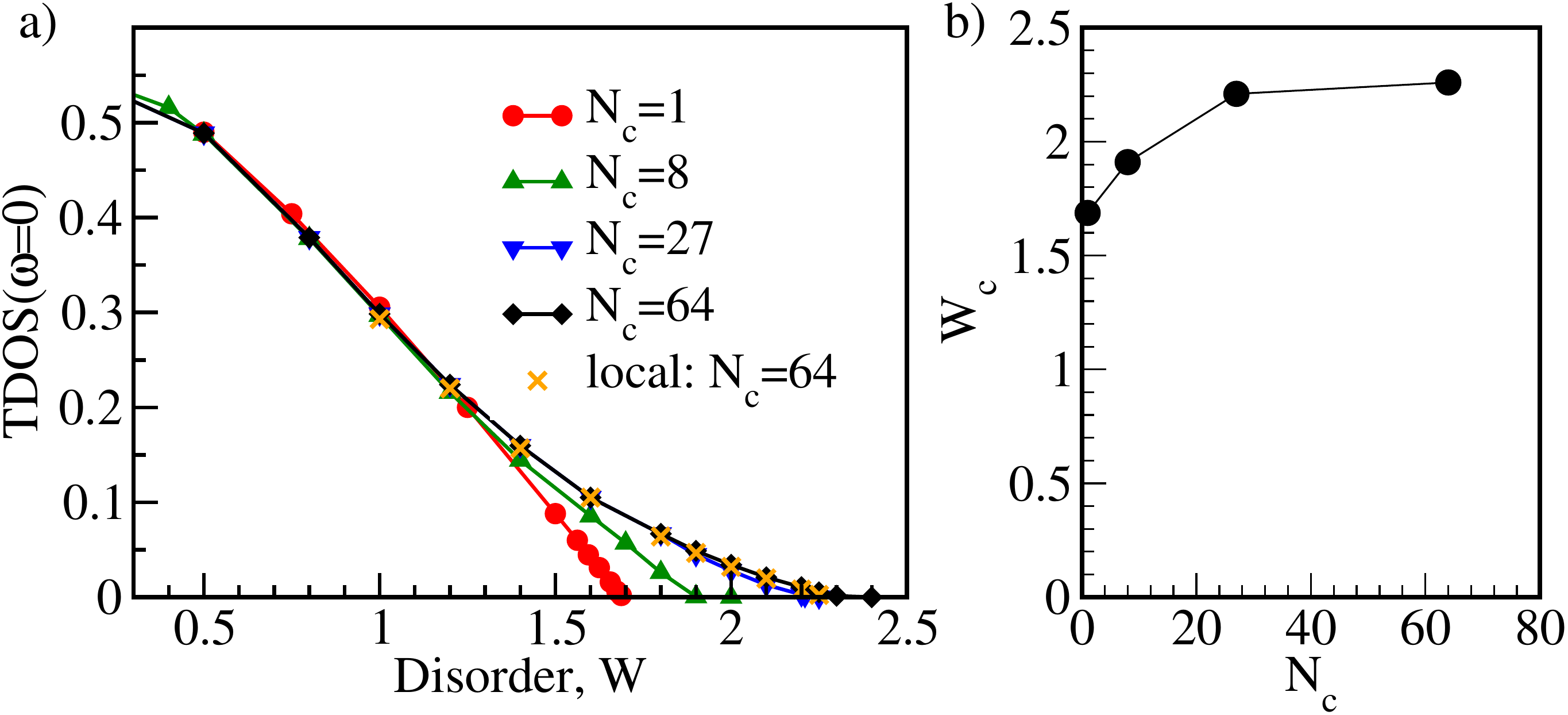}
\caption[]{(Left a) panel): The $TDOS(\omega=0)$ (obtained using ansatz of Eq.~\ref{eq: ansatz}) vs disorder strength $W$ for different cluster sizes $N_c=1, 8, 27, 64$. For $N_c=64$ cluster, we also show the data for $TDOS(w=0)$ calculated using local ansatz of Eq.~\ref{eq:local-ansatz} (shown by orange crosses). (Right b) panel: The critical disorder strength of the Anderson transition $W_c$ as function of $N_c$. The critical disorder strength $W_c$ is determined from $TDOS(w=0)=0$ data of the left panel.}
\label{fig:Wc-box}
\end{figure} 

Next, we consider the evolution of the critical disorder strength $W_c$ of the Anderson localization as a function of the cluster size $N_c$. The critical disorder strength $W_c$ is obtained by calculating the TDOS at the band center TDOS($w$=0) as a function of disorder strength $W$. The $W_c$ is then defined by vanishing TDOS($w$=0)=0. We have done such analysis for several cluster sizes $N_c=1, 8, 27, 64$ on a cubic 3D Anderson model lattice. Our results are shown in Fig.~\ref{fig:Wc-box} (panel a)), where we plot the $TDOS(w=0)$ at the band center as a function of disorder strength $W$. The TDOS($w$=0) decreases with increasing disorder strength $W$, and approaches zero at $W_c$. Performing a careful analysis for different clusters $N_c$, we demonstrate that the critical disorder strength $W_c$ converges quickly with the cluster size $N_c$ (panel b) of Fig.~\ref{fig:Wc-box}). These data also highlight the importance of going beyond the single-site approximation when describing the critical behavior of the Anderson transition. For $N_c=1$ (which corresponds to the local TMT approximation case) the $W_c \approx 1.675$, and increases gradually to the converged value of $W_c \approx 2.25$ as cluster size $N_c$ gets larger. These results are in a good agreement with the $W_c$ reported in the literature. While TMDCA slightly overestimates the $W_c$ as compared to exact results, the advantage of our method  is that it can be applied for the interacting and realistic systems. Finally, for $N_c=64$, we also show the results obtained with the local ansatz for the TDOS of Eq.~\ref{eq:local-ansatz} (the corresponding data are shown by the orange crosses in Fig.~\ref{fig:Wc-box}). The local TDOS data fall on the top of the $TDOS(w=0)$ obtained using full $K$ dependence. This indicates the critical disorder strength $W_c$ of Anderson localization can be very well captured by simplified ansatz of Eq.~\ref{eq:local-ansatz}, which should be an important simplification when applying the TMDCA for more realistic models.

\subsection{Binary alloy disorder distribution}

\begin{figure}[htb]
\includegraphics[width=1\textwidth]{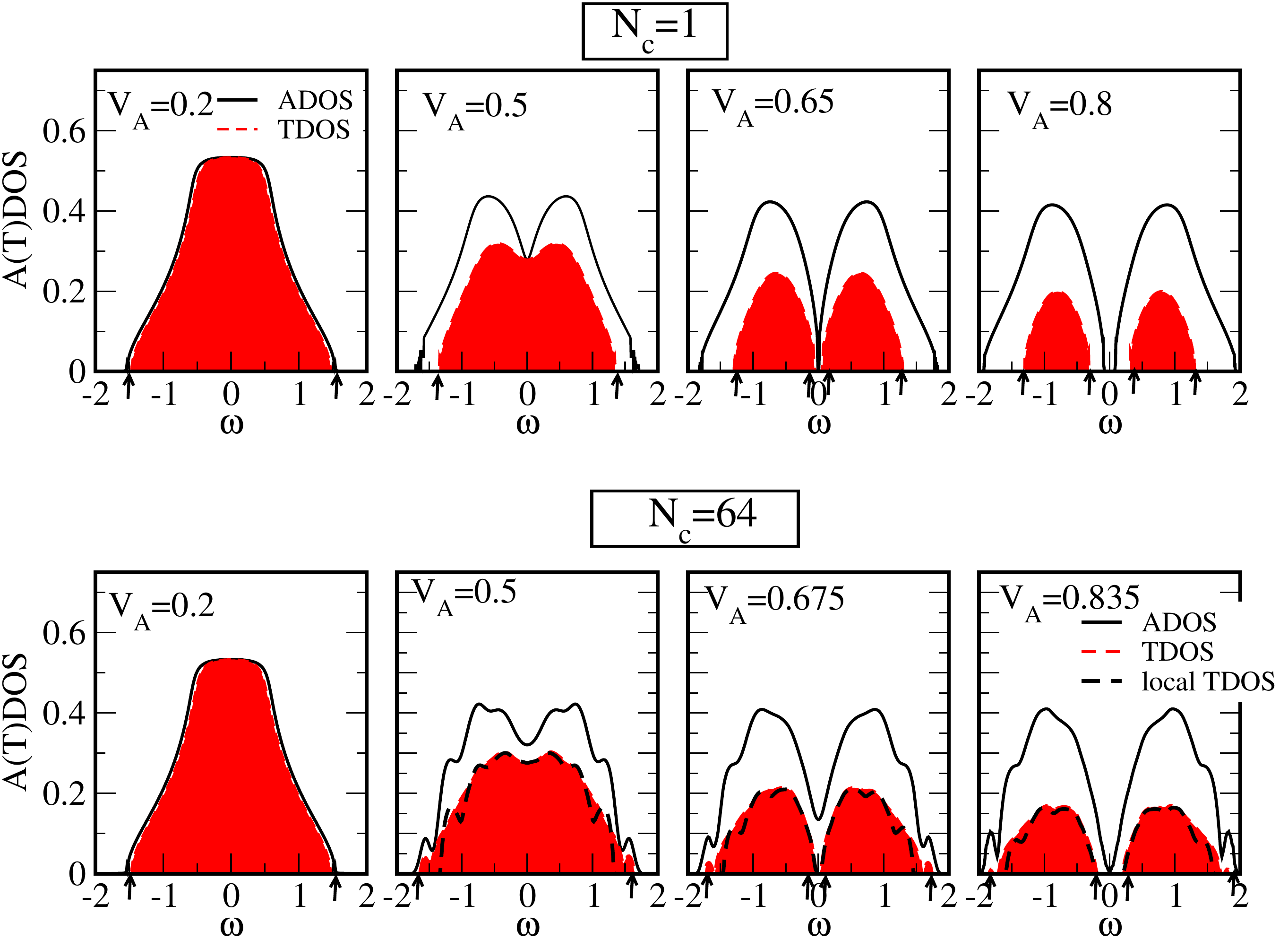}
\caption[]{(Top panel): $N_c=1$ CPA and TMT results for the ADOS($w$) and TDOS($w$) at different disorder strength values $V_A=0.2, 0.5, 0.65., 0.8$. (Bottom panel): $N_c=64$ DCA and TMDCA results for the ADOS($w$) and TDOS($w$) as function of increasing disorder $V_A=0.2, 0.5, 0.675, 0.835.$ The $N_=64$ TMDCA data for the TDOS($w$) (shaded region) are obtained using ansatz of Eq.~\ref{eq: ansatz}, the local TDOS($w$) curves (dashed lines) are obtained using a simplified local ansatz of Eq.~\ref{eq:local-ansatz}. Other parameters: $c_a=0.5$}
\label{fig:dos-binary}
\end{figure}

To further explore the application of our TMDCA approach, we now consider its implementation for the binary alloy systems. 
First, in Fig.~\ref{fig:dos-binary}, we show the results for the ADOS($\omega$) and TDOS($\omega$) at different disorder strength $V_A$. To highlight the significance of the non-local spatial effects, we present the data for the local $N_c=1$ (TMT) and the non-local $N_c=64$ TMDCA case. For binary alloy systems, when increasing the disorder strength $V_A$, the system undergoes two phase transitions, i.e., the Anderson transition which is detected by vanishing $TDOS(w=0$) at the band center, and the band-gap opening insulator transition detected by vanishing $ADOS(w=0$) at the Fermi level. For small disorder strength ($V_A=0,2$), both the ADOS and the TDOS are practically the same. As disorder strength increases, the band gap opens in both the ADOS and the TDOS at the Fermi level. For the local $N_c=1$ case, the Anderson localization of the states at the Fermi level and the band splitting transition occur at almost the same disorder strength, i.e., the TDOS$(w=0)=0$ at $V_c^{typ}\approx 0.6275$, and ADOS$(w=0)=0$ at $V_c^{ave}\approx 0.635$. As disorder strength $V_A$ increases, the TDOS becomes significantly smaller and narrower than the ADOS. Regions where $ADOS(w)$ remains finite but $TDOS(w)$ is zero, indicate the Anderson localized states, separated by the mobility edges (marked by vertical arrows). Comparing the $N_c=1$ and finite cluster $N_c=64$ results, we find that the non-local spatial correlations, which are included in the TMDCA scheme, introduce a noticeable difference in the localization behavior. In particular, the non-local effects are responsible for the finite structures in the $ADOS(w)$, which are completely smoothed out in the local $N_c=1$ case. Moreover, we find that the Anderson transition at the Fermi energy with $TDOS(w=0)$ occurs much faster than the gap opening in the $ADOS(w)$. $N_c=64$ results also indicate that the mobility edge trajectories are wider than in $N_c=1$ case, i.e., the local TMT scheme underestimates the extended states region \cite{c_ekuma_14a, Terletska_etal_2018}. We also show the results (black dashed lines) for $N_c=64$ local $TDOS(w)$ obtained using a simplified ansatz of Eq.~\ref{eq:local-ansatz}. As in the described above box distribution case, we see that the major contribution to the $TDOS(w)$ is coming from the geometrically averaged local TDOS factor. The full $K$- dependence in the ansatz of Eq.~\ref{eq: ansatz} seems to be mostly relevant for capturing the higher frequency mobility edge behavior.

\begin{figure}[htb]
\includegraphics[width=1\textwidth]{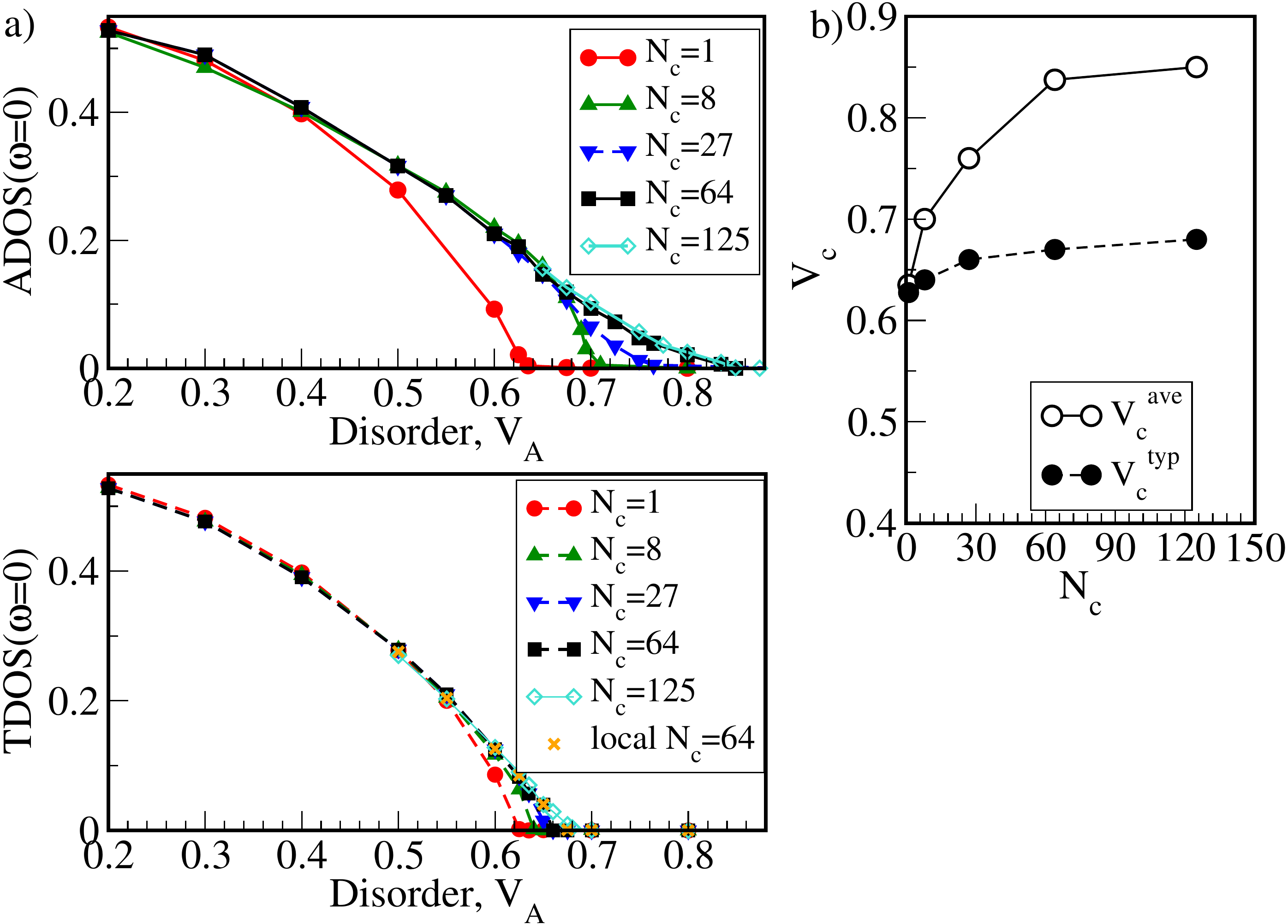}
\caption[]
{(Right a) panel). Top: ADOS$(w=0)$ as function of disorder strength $V_A$ for different cluster sizes $N_c=1, 8, 27, 64, 125$. Bottom: TDOS$(w=0)$ as function of disorder strength $V_A$ for different cluster sizes $N_c=1, 8, 27, 64, 125$.
(Left b) panel): The critical disorder strength of the Anderson transition $V_c^{typ}$ and of the band-splitting transition $V_c^{ave}$ as function of the cluster size $N_c$. $V_c$ are determined from the corresponding TDOS$(\omega= 0) = 0$ and ADOS$(w=0)=0$ data of the left panel. Other parameters: $c_a=0.5$}
\label{fig:Wc-binary}
\end{figure}

To explore the cluster size dependence of the critical disorder strength $V_c$ for the corresponding transitions in the binary alloy model, in Fig.~\ref{fig:Wc-binary} we plot the ADOS$(w=0)$ (a) panel, top graph) and the TDOS$(w=0)$ (a) panel, bottom graph) as a function of increasing disorder strength $V_A$ at different cluster sizes $N_c=1, 8, 27, 64, 125$. The critical value for the Anderson localization transition $V_c^{typ}$ is then extracted from the vanishing $TDOS(w=0)$, and zero in $ADOS(w=0)$ determined the critical disorder strength $V_c^{ave}$ for the band-gap opening transition. For the $N_c=1$, the Anderson localization and the the band-splitting transition occur almost simultaneously. However, at $N_c>1$, the Anderson localization  clearly proceeds the band-gap opening, and occurs at smaller values of $V_c$. The difference in cluster size $N_c$ convergence is better observed in Fig.~\ref{fig:Wc-binary} (b) panel), where we plot the critical disorder strength $V_c^{typ}$ and $V_c^{ave}$ (extracted from the corresponding data on the left panels) as function of the cluster size $N_c$. We observe that $V_c^{typ}$ converges rather quickly with increasing cluster size $N_c$, while there is significant $N_c$ dependence of $V_c^{ave}$ for the band-splitting transition implying the importance of strong non-local spatial correlations.


\section{Conclusion}
\label{sec: conclusion}
We use the developed finite cluster typical medium TMDCA approach to study the Anderson localization in three dimensions for the box and binary alloy disorder distributions. By performing a careful systematic cluster size analysis, we demonstrate that TMDCA presents a successful and numerically manageable effective medium embedding approach for the Anderson localization. We show that non-local correlations are significant for the proper analysis of the Anderson transition, and hence the importance of employing beyond the local single site approximations. Using the typical density of states as the order parameter for Anderson transition, we obtained the cluster size converged critical disorder strengths for the transition $W_c\approx 2.25$ (box distribution) and $V_c\approx 0.675$ (binary alloy case), which is in good agreement with the known results in the literature. Finite cluster $N_c$ values for the critical disorder strength $W_c$ are of noticeable improvement over the local single-site TMT results with $W_c\approx 1.68$, and $V_c\approx 0.6275$ for the box and binary distributions, respectively. We have also demonstrated the importance of the non-local correlations in capturing the spectral properties of the disordered systems. The application of the TMDCA to more realistic and complex models is often faced with the challenge of constructing a proper ansatz for the calculation of the geometrically averaging Green's function in the self-consistency numerical loop. Our results show that the geometrically averaged local density of state factor (in the expression for the typical medium Green's function ansatz of Eq. ~\ref{eq: ansatz}) is important for capturing properly the localization at the band center and the critical disorder strength. While the non-local momentum $K$- dependent part in the ansatz plays felicitates a faster cluster size convergence of the localized states at the edges. The performed analysis presented in this work will allow for more effective application of the developed TMDCA method to more complex disorder models, and provide a better understanding of the role of non-local spatial correlations in the disordered systems. 

\section{Declaration of competing interest}
The authors declare that they have no known competing financial interests or personal relationships that could have appeared to influence the work reported in this paper.

\section{Acknowledgement}
HT has been supported by NSF OAC-1931367 and NSF DMR-1944974 grants. KMT is  partially supported by NSF DMR-1728457 and NSF OAC-1931445. YW is partially supported by NSF OAC-1931525. The work of ME has been supported by U.S. Department of Energy,  Office  of  Science,  Basic  Energy  Sciences,  Material Sciences and Engineering Division and it used resources of the Oak Ridge Leadership Computing Facility,which is a DOE Office of Science User Facility supported under Contract DE-AC05-00OR22725.  LC gratefully acknowledges the financial support offered by the Augsburg Center for Innovative Technologies and by the Deutsche Forschungsgemeinschaft (DFG, German Research Foundation) - Projektnummer 107745057 - TRR 80/F6. JM is supported by the U.S. Department of Energy, Office of Science, Office of Basic Energy Sciences under Award Number DE-SC0017861.

\bibliographystyle{elsarticle-num-names}
\bibliography{Proceedings_localisation_Terletska}

\end{document}